\documentclass[aps,preprint,pre,superscriptaddress,runinaddress,showpacs]{revtex4}
\usepackage{amssymb}
\usepackage{amsmath}
\usepackage{graphicx}
\usepackage{bm}

\begin{document}

\title{Choosing Hydrodynamic Fields}
\author{James W. Dufty}
\affiliation{Department of Physics, University of Florida, Gainesville, FL 32611, USA}
\author{J. Javier Brey}
\affiliation{F\'{\i}sica Te\'{o}rica, Universidad de Sevilla, Apartado de Correos 1065,
E-41080 Sevilla, Spain}
\date{\today }

\begin{abstract}
Continuum mechanics (e.g., hydrodynamics, elasticity theory) is based on the
assumption that a small set of fields provides a closed description on large
space and time scales. Conditions governing the choice for these fields are
discussed in the context of granular fluids and multi-component fluids. In
the first case, the relevance of temperature or energy as a hydrodynamic
field is justified. For mixtures, the use of a total temperature and single
flow velocity is compared with the use of multiple species temperatures and
velocities.
\end{abstract}

\maketitle

\section{Introduction}

\label{sec1}Continuum mechanics (e.g., hydrodynamics, elasticity theory)
provides the macroscopic description for a wide class of physical systems
and states in terms of a few space-time fields. The derivation and
justification of the equations for these fields from the underlying
Newtonian or quantum mechanics is a well-studied and open problem of
non-equilibrium statistical mechanics \cite{McLennan,Zwanzig,Grabert}. The
central conceptual and practical problem is the reduction of the many
degrees of freedom in the microscopic description to a closed description in
terms of a few chosen macroscopic fields. Qualitatively, this is understood
in terms of restrictions on the conditions for applicability of continuum
mechanics. In particular, it is expected that such a reduced description can
apply only for slowly varying fields that dominate the description on
sufficiently long time scales. Under such conditions, all other fast degrees
of freedom have decayed to zero leaving the possibility for a reduced
description.

The choice of fields is therefore an important first step in any derivation
of continuum mechanics. This in turn requires understanding relevant time
scales active in the system. For a simple molecular fluid, there is a time
independent uniform state, equilibrium, which is the reference state for the
dynamics. The longest time scale for approach to equilibrium is set by the
relative size of the spatial deviations of the relevant macroscopic fields $%
y $, $\mu \sim \Delta y/y_{e}$, where $\Delta y\sim y\left( \mathbf{r}+%
\mathbf{r}_{0},t\right) -y\left( \mathbf{r},t\right) $, $y_{e}$ is the
equilibrium value of $y$, and $\mathbf{r}_{0}$ is a distance of the order of
the mean free path. In the following, $\mu $ is referred to as the \emph{%
uniformity parameter}. The local conserved fields of number, energy, and
momentum densities obey balance equations expressing their time derivatives
as gradients of fluxes. These derivatives are therefore proportional to the
uniformity parameter, identifying a time scale that diverges as the system
approaches equilibrium. A necessary condition for choosing a set of fields,
therefore, is that they should include all local conserved densities. In the
case of broken symmetries (e.g. solids, liquid crystals, superfluids) there
are additional non-conserved fields whose time derivatives are proportional
to the uniformity parameter, and therefore must also be included in the set
of chosen fields \cite{Forster,Martin}. Finally, there are systems with
internal processes on long time scales (polymers, plasmas) that remain
finite in the uniform limit. Their associated fields can be included in the
chosen set if these time scales are long compared to all other non-conserved
excitations.

In the following, attention will be restricted to the fluid phase, for which
the macroscopic description is hydrodynamics. Also it is assumed that there
are no slow excitations associated with a broken symmetry. For reasons just
noted, the hydrodynamic fields must include all conserved densities as a
necessary condition to be complete on the longest time scale. The question
addressed here is whether this minimal set can be extended to include
additional fields, and if so, whether it is appropriate to do it. This issue
has been discussed extensively in the context of extended irreversible
thermodynamics (EIT) \cite{Jou}. In EIT, additional fields, such as the
fluxes in the balance equations for the conserved densities are included as
independent hydrodynamic fields. One motivation there is to resolve
paradoxes of Navier-Stokes hydrodynamics, usually associated with a misuse
of those equations for conditions where they do not apply (e.g., initial
slip, boundary layers, shock fronts). Another approach using additional
fields is the moment method of Grad \cite{Lutsko}. Here, motivated by
current applications of hydrodynamics to granular fluids \cite{General},
related but different issues are addressed.

Section \ref{sec2} outlines the basic ingredients for a derivation of
hydrodynamic equations near a state of uniformity, from the underlying
equations of nonequilibrium statistical mechanics \cite{Dufty}. It relies on
exact balance equations for the chosen fields, obtained from the Liouville
equation, and on a special ``normal'' solution to that equation.
Construction of a approximate normal solution is obtained by expansion in
the uniformity parameter, leading to Navier-Stokes level hydrodynamics. The
derivation does not make any explicit limitation on the chosen fields, other
than the existence of a uniform reference state expressible in terms of
those fields. However, it does not provide the context under which such a
solution is applicable. It is argued that a necessary condition must be
sufficiently long time scales.

In Sec.\,  \ref{sec3} the example of a simple granular fluid is considered.
The uniform state in this case is not equilibrium, but one for which the
temperature decreases monotonically due to inelastic collisions. The most
significant differences from a molecular fluid are non-conservation of
energy and a uniform state that is time dependent. The number and momentum
densities are still conserved, so they are included in the set of
hydrodynamic fields. The question of including the non-conserved energy
density (or, equivalently, the granular temperature) is discussed. It is
concluded that the minimal set of hydrodynamic fields must include the
temperature.

In Sec.\, \ref{sec4} binary mixtures are addressed, for both molecular and
granular fluids. Two sets of fields are considered for hydrodynamic
descriptions. One is the conserved number densities, flow velocity
associated with the conserved total momentum, and the temperature associated
with the total energy. The second one is an expanded set consisting of the
conserved number densities, the species flow velocities associated with the
non-conserved species momenta, and the species temperatures. It is argued
that the second description, while more detailed, has no predictive value
beyond the simpler hydrodynamic description on the relevant large space and
time scales. The presentation is summarized in the last section with some
concluding remarks.

\section{Normal solutions, constitutive equations, and hydrodynamics}

\label{sec2}The most general notion of a hydrodynamic description is a
closed set of equations for a set of  hydrodynamic fields, denoted by $\left\{
y_{\alpha }\right\} $. The terminology ``closed'' means that they obey a set
of equations of the form
\begin{equation}
\partial _{t}y_{\alpha }\left( \mathbf{r},t\right) =N_{\alpha }(\mathbf{r}
,t | \left\{ y_{\beta} \right\} ),  \label{2.1}
\end{equation}
where the space, ${\bm r}$, and time, $t$, dependence of the right side is entirely determined
from the fields themselves. In general, $N_{\alpha }(\mathbf{r},t | \left\{
y_{\beta }\right\} )$ is a nonlinear functional of the fields. The
derivation of such equations procedes in two steps. First, balance equations
for the fields are obtained directly from the Liouville equation,
\begin{equation}
\partial _{t}y_{\alpha }\left( \mathbf{r},t\right) +\nabla \cdot \mathbf{j}%
_{\alpha }\left( \mathbf{r},t\right) =s_{\alpha }(\mathbf{r},t).  \label{2.2}
\end{equation}%
Here, the fluxes $\mathbf{j}_{\alpha }\left( \mathbf{r},t\right) $ and
sources $s_{\alpha }(\mathbf{r},t) $ are linear, time-independent
functionals of the solution to the Liouville equation. In general, these
functionals cannot be characterized entirely by the hydrodynamic fields, so
although (\ref{2.2}) are exact equations they do not constitute a closed set
of equations for the fields in terms of themselves. This closure occurs for
the special conditions of a ``normal'' state, defined as a solution to the
Liouville equation of the form
\begin{equation}
\rho _{n}\left( \Gamma ,t\right) =\rho _{n}\left( \Gamma | \left\{
y_{\alpha} \right\} \right) =\rho _{n}\left( \left\{ \mathbf{q}_{ij}\right\}
,\left\{ \mathbf{v}_{i}\right\} ,\left\{ y_{\alpha} \left( \mathbf{q}%
_{i},t\right) \right\} \right),  \label{2.3}
\end{equation}
where ${\bm q}_{i}$ and ${\bm v}_{i}$ are the position and velocity of
particle $i$, and ${\bm q}_{ij} \equiv {\bm q}_{i}-{\bm q}_{j}$. For such
solutions, all time dependence and the breaking of translational invariance
occur only through the hydrodynamic fields. The notation in the second
equality indicates that the \emph{functional} of the fields $\rho _{n}\left(
\Gamma \mid \left\{ y_{\alpha} \right\} \right) $ is equivalent to a \emph{\
function} of the fields at the particle positions of the specified $\Gamma $
point. For a normal solution, the definitions of $\left\{ \mathbf{j}_{\alpha
}\left( \mathbf{r},t\right) \right\} $ and $\left\{ s_{\alpha }(\mathbf{r}%
,t)\right\} $ as linear functionals of this solution lead to ``constitutive
equations''
\begin{equation}
\mathbf{j}_{\alpha }\left( \mathbf{r},t\right) =\mathbf{j}_{\alpha }(\mathbf{%
\ r,}t | \left\{ y_{\beta} \right\} )\mathbf{,\hspace{0.25in}}s_{\alpha }(%
\mathbf{r} ,t)=s_{\alpha }(\mathbf{r,}t | \left\{ y_{\beta} \right\} ).
\label{2.4}
\end{equation}
With the fluxes and sources now determined by the fields, the balance
equations (\ref{2.2}) become closed hydrodynamic equations of the form (\ref%
{2.1}).

Since the balance equations are exact consequences of the microscopic
dynamics, questions about the existence of a hydrodynamic description turn
on the necessary conditions for a normal solution (e.g., appropriate choice
for the set of fields $\left\{ y_{\alpha}\right\} $, restrictions on the space and
time scale). Substitution of (\ref{2.3}) into the Liouville equation gives
the equation a normal solution must satisfy for a given choice of fields $%
\left\{ y_{\alpha} \right\} $,
\begin{equation}
\int d\mathbf{r}\, \frac{\delta \rho _{n}}{\delta y_{\alpha }\left( \mathbf{r}
,t\right) }N_{\alpha }(\mathbf{r},t | \left\{ y_{\beta} \right\} )+L \rho _{n}=0,
\label{2.5}
\end{equation}%
where summation over repeated indices is implicit, $L$ is the Liouville
operator of the system, and $N_{\alpha }(\mathbf{r},t | \left\{ y_{\beta
}\right\} )$ is the nonlinear functional in the hydrodynamic equations (\ref%
{2.1}), defined in terms of the fluxes and sources. The latter in turn must
be determined self-consistently in the form of constitutive equations (\ref%
{2.4}), from the solution to (\ref{2.5}). The resulting solution is
implicit, being a function of the phase point $\Gamma $ and a functional of
the unknown fields. The space and time dependence of these fields is then
obtained from a solution to the hydrodynamic equations for given initial and
boundary conditions, completing specification of the normal solution.

Determination of a normal solution in this way is quite difficult, but its
approximate construction can be carried out explicitly for states with small
uniformity parameter $\mu $ (described above), by expansion in this
parameter. This construction of the normal solution by expansion around the
corresponding homogeneous solution will be referred to as the Chapman-Enskog
method \cite{Chapman,Origins}. This presumes the existence and determination
of a reference homogeneous solution $\rho _{h}$ to Eq.\thinspace\ (\ref{2.5}%
),
\begin{equation}
\frac{\partial \rho _{h}}{\partial y_{\alpha }\left( t\right) }s_{\alpha
}(\left\{ y_{\beta }\left( t\right) \right\} )+L\rho _{h}=0,  \label{2.6}
\end{equation}%
together with the homogeneous hydrodynamic equations%
\begin{equation}
\partial _{t}y_{\alpha }\left( t\right) =s_{\alpha }(\left\{ y_{\beta
}\left( t\right) \right\} ).  \label{2.7}
\end{equation}%
The familiar example is a one component atomic fluid near spatially
homogeneous equilibrium. If the fields chosen are all associated with local
conserved densities, then the sources $s_{\alpha}$ vanish and (\ref{2.6}) becomes the
condition for a stationary equilibrium state. More generally, if the fields
have sources, the homogeneous solution depends on the time dependence of
these fields generated by the sources. Effects of small spatial gradients
are described by returning to (\ref{2.5}) and expanding the solution about $%
\rho _{h}$ to linear order in $\mu $,
\begin{equation}
\rho _{n}\left( \Gamma \mid \left\{ y_{\alpha }\right\} \right) =\rho
_{h}\left( \Gamma ;\left\{ y_{\alpha }\right\} \right) +\Delta _{\alpha
}(\Gamma ;\left\{ y_{\beta }\left( \mathbf{r},t\right) \right\} )\cdot \mu
\nabla y_{\alpha }(\mathbf{r},t)+\ldots  \label{2.8}
\end{equation}%
The dots indicate terms of higher order in $\mu $, and (\ref{2.5}) provides
a linear inhomogeneous equation for determination of $\Delta _{\alpha }.$
With this result, the constitutive equations also are determined to first
order in the gradients, providing $N_{\alpha }$ and the hydrodynamic
equations to this order.

If the fields are chosen to be the local conserved densities of mass,
energy, and momentum, then the sources $s_{\alpha }(\mathbf{r} ,t) $ all
vanish, and the fluxes $\mathbf{j}_{\alpha }( \mathbf{r},t) $ are
proportional to the uniformity parameter. Construction of the normal
solution by the Chapman-Enskog method determines the time derivatives to be
proportional to the uniformity parameter as well. The leading approximation
yields the perfect fluid Euler equations, and the next order gives the usual
Navier-Stokes equations. The derivation provides their context for
applicability as well in this case: states must be near uniform equilibrium,
with small fractional variations in the fields over a mean free path, and
space and time scales large compared to the mean free path and mean free
time, respectively.

It would appear that this approach is applicable for any choice of fields,
leading to many forms for the macroscopic description of a system. However,
the conditions under which a normal solution could be expected must be
considered further. Consider first a normal fluid with elastic collisions in
an initial non-equilibrium state with specified hydrodynamic fields $\left\{
y_{\alpha }\left( \mathbf{r},t=0\right) \right\} $, whose values vary
smoothly across the system. In each small region of dimension larger than a
mean free path, the phase space density $\rho (\Gamma ,t)$ approaches a
Gibbs density characterized by the hydrodynamic fields at its central point $%
\mathbf{r}$. Subsequently, exchange of mass, energy, and momentum tends to
equilibrate these fields to uniform values (or to steady values if the
system is driven). The first stage, approach to a universal form for the
velocity distribution, occurs after a few collisions. This establishes the
normal form of the solution, where the hydrodynamic fields and their
gradients characterize the state. Deviations from the equilibrium
distribution are due to fluxes of mass, momentum, and energy across the
cells. These fluxes are proportional to the differences in values of the
fields (i.e., to their spatial gradients). The second stage is the slower
evolution of the distribution through the changing values of the fields,
according to the ``usual'' Navier-Stokes hydrodynamic equations. In this
scenario, there is a characteristic microscopic time scale (mean free time)
and a characteristic microscopic space scale (mean free path) that set the
limits for validity of the normal solution. Beyond these limits, the
dynamics is complex and not captured entirely by that of the hydrodynamic
fields.

Extending the hydrodynamic fields to include non-conserved fields
necessarily introduces additional time scales in the hydrodynamic solutions,
due to the sources $s_{\alpha }$ associated with the new fields. Thus the
hydrodynamics of the second stage now has variations on the same time scale
as the first stage. Furthermore, the additional fields give only a partial
description of the short time scale dynamics, and therefore do not extend
the domain of applicability for the hydrodynamic description. The
qualitative initial stage described above for a general initial preparation
of the system is still required for any reduction in the large number of
degrees of freedom active in the system to decay. The hydrodynamics can be
closed (Markovian) only after such a reduction. Even during the second
stage, only the slower dynamical processess can be described by
hydrodynamics, as there are continual fluctuations of the fast degrees of
freedom that would prohibit isolation of the corresponding fast hydrodynamic
modes. In summary, the choice of fields for hydrodynamic equations is
somewhat arbitrary, but their dominance on some time scale is a central
property for utility as a macroscopic description of the system. Predictions
on shorter time scales are difficult to correlate with measurements, and
potentially obscuring the simplicity of the larger time scale dynamics. On
the common long time scale required for validity, it would appear that the
normal solutions should be equivalent, with the more detailed fields
becoming functionals of the simpler fields and being \textquotedblleft
slaved\textquotedblright\ to them.

Exceptions to this conclusion occur when the uniform reference state is not
the equilibrium state, and may have an inherent dynamics of its own. This is
the case for granular fluids. In the following, the hydrodynamic
descriptions for two examples using \ non-conserved fields are discussed: a
simple one-component granular fluid for which there is an energy source, and
binary mixtures where non-conserved fields of individual species may be of
interest. Attention is limited to states near a corresponding universal
homogeneous state (homogeneous cooling or equilibrium state) for simplicity.
The question addressed is \textquotedblleft what are the appropriate
Navier-Stokes hydrodynamic equations for these more complex conditions, and
under what conditions do they apply?\textquotedblright .

\section{Granular fluid}

\label{sec3}A one-component granular fluid consists of \textquotedblleft
particles\textquotedblright\ (grains) of mass $m$ with short ranged
collisions that conserve number and momentum, but not energy. By analogy
with a molecular fluid it is reasonable to expect a hydrodynamic description
in terms of the local number, momentum, and energy densities. It is usual to
replace the momentum by its associated flow velocity $\mathbf{u}$, and the
energy by a granular temperature $T$. This is just a change of variables.
The exact macroscopic balance equations for $n,\mathbf{u}$, and $T$ are \cite%
{BDyS97}
\begin{equation}
D_{t}n+n\nabla \cdot \mathbf{u}=0,  \label{3.1}
\end{equation}%
\begin{equation}
D_{t}\mathbf{u}+\frac{1}{\rho}\nabla \cdot \mathsf{P}=0,
\label{3.2}
\end{equation}%
\begin{equation}
D_{t}T+\frac{2}{3n}\left( \nabla \cdot \mathbf{q}+\mathsf{P}:\nabla \mathbf{u%
}\right) =-\zeta T.  \label{3.3}
\end{equation}%
where $D_{t}\equiv \partial _{t}+\mathbf{u}\cdot \nabla $ is the
material derivative, and $\rho =nm$. For simplicity, the particles
have been represented as hard spheres for which the temperature is
defined in terms of the energy density $e$ as $e=3nT/2+\rho u^2/2$.
The heat flux $\mathbf{q}$ and pressure
tensor $\mathsf{P}$ are defined as the energy and momentum fluxes
in the local rest frame of a fluid element moving with
velocity $\mathbf{u}$. They are related with the energy and momentum fluxes
in the laboratory frame, $\mathbf{s}$ and $\mathsf{T}$, by
\begin{equation}
\mathbf{q=s-}e\mathbf{u-u}\cdot \mathsf{P\mathbf{,\hspace{0.25in}}P=T}-\rho
\mathbf{uu.}  \label{3.3a}
\end{equation}%
These balance equations have the same form as for a molecular fluid, except
for the presence of a source $-\zeta T$ in the temperature equation,
resulting from the inelasticity of collisions among the particles. The
parameter $\zeta $ will be referred to as the cooling rate. Application of
the Chapman-Enskog method for a normal solution to first order in the
uniformity parameter leads to constitutive equations for the heat and
momentum fluxes, $\mathbf{q}$ and $\mathsf{P}$, similar to Fourier's law and
Newton's viscosity law \cite{DByB08}
\begin{equation}
\mathbf{q=}-\lambda \nabla T-\mu \nabla n,\hspace{0.25in}P_{ij}=p\delta
_{ij}-\eta \left( \nabla _{i}u_{j}+\nabla _{j}u_{i}-\frac{2}{3}\delta
_{ij}\nabla \cdot \mathbf{u}\right) -\kappa \delta _{ij}\nabla \cdot \mathbf{%
u}.  \label{3.4}
\end{equation}%
In addition, the constitutive equation for the cooling rate $\zeta $ up to first order in the gradients
is given by%
\begin{equation}
\zeta =\zeta _{0}+\zeta _{1}\nabla \cdot \mathbf{u}.  \label{3.5}
\end{equation}%
The pressure $p$ and the transport coefficients in these expressions, $\lambda $, $\mu $, $\eta $%
, $\kappa $, $\zeta _{0}$, and $\zeta _{1}$, are also determined from the
normal solution as functions of $n$ and $T$.

Equations (\ref{3.1})-(\ref{3.5}) constitute the Navier-Stokes order
hydrodynamic equations for a simple granular fluid. In contrast to a
molecular fluid, solutions to the granular hydrodynamic equations have two
time scales, one set by the uniformity parameter and another set by the
energy loss rate $\tau \sim $ $\zeta _{0}^{-1}$. The former becomes large as
the system approaches uniformity, while the latter remains finite in uniform
systems. From the discussion of the last section, in order for the
hydrodynamic description to dominate the complex microdynamics, both of
these time scales should be large compared to the mean free time. In
practice this appears to be the case, at least if the degree of inelasticity
is not too large (since $\zeta _{0}$ vanishes in the elastic limit).

As the system approaches uniformity, the two hydrodynamic time scales become
very different and the question arises of whether the temperature is still a
relevant hydrodynamic field. It is a \textquotedblleft fast
variable\textquotedblright\ on the time scale for the spatial variations, so
perhaps there is a simpler description in terms of $n$ and $\mathbf{u}$
alone. To see that this is not the case consider Eq.\ (\ref{2.6}) for the
reference homogeneous normal solution,
\begin{equation}
\left( -\zeta T\frac{\partial }{\partial T}+L\right) \rho _{h}=0.
\label{3.6}
\end{equation}%
To exclude the temperature as a relevant variable, $\rho _{h}$ should be
independent of the temperature, i.e. $T$ should represent a transient
dynamics that vanishes or becomes constant on a sufficiently long time
scale. Then (\ref{3.6}) would become $L\rho _{h}=0$ for the reference
homogeneous normal solution on this time scale. However, there is no
solution to this equation for finite inelasticity. This can be easily seen
by considering the average rate of change in the energy $E$,
\begin{equation}
\partial_{t} \left\langle E\right\rangle =-\int d\Gamma\, EL\rho _{h}.  \label{3.6a}
\end{equation}%
This can never vanish since the collisional cooling continues as the system
evolves. Thus the temperature is an inherent property of all solutions and
cannot be neglected on any time scale. Although $\zeta _{0}$ characterizes a
rapid variation relative to the spatial time scale, it persists as a
modulation of that latter slow scale. The reference uniform state has a
dynamics (cooling) that is inherited by spatial deviations from that state
and the system continues to cool rapidly even after it has \textquotedblleft
aged\textquotedblright\ an arbitrarily long time.

It might be argued that for special boundary conditions the hydrodynamic
equations would support a solution with stationary temperature, such that
variations of the temperature are indeed transient and rapidly approaching
its stationary value. Subsequently, the residual relaxation of spatial
variation might be described by $n$ and $\mathbf{u}$ alone, and only
parameterized by a constant temperature. While such a contrived description
might be possible on a case-by-case basis, this does not constitute an
alternative hydrodynamics. A hydrodynamic description is a set of closed
equations for chosen fields that describes a class of states, e.g.
Navier-Stokes equations for states near homogeneity, whose form is universal
and independent of specific initial or boundary conditions.

The granular fluid with the source $-\zeta T$ can be contrasted with a
molecular fluid including non-conserved fields. In the latter case, the
sources vanish for the homogeneous (equilibrium) state so that solutions to $%
L\rho _{e}=0$ do exist. There is then no restriction on the removal of the
non-conserved fields in that case, for a simpler hydrodynamics. In contrast,
the granular source discussed here does not vanish for its homogeneous state
and hence the dynamics due to this source must be included in any set of
hydrodynamic fields.

In summary, the Navier-Stokes hydrodynamics for a simple granular fluid is
given in terms of $n,\mathbf{u}$, and $T$ \ just as for a molecular fluid.
Due to the inelasticity of collisions, there is a new time scale set by the
cooling rate that does not increase as the system approaches spatial
homogeneity. As for molecular fluids, the normal solution and hydrodynamics
applies for times long compared to the mean free time when the transient
dynamics of microscopic degrees of freedom has decayed and only that of the
fields remains. However, the granular hydrodynamics retains a modulation of
the spatial relaxation due to continual cooling of the system.

\section{Binary mixtures}

\label{sec4}

Consider next a granular or molecular fluid composed of two different
species (e.g., different mass, size or collision law). If there are no
reactions or fragmentation, the number of particles of each species is
conserved and their local densities, $n_{i}$, are appropriate independent
hydrodynamic fields. Other properties of interest might be the species
temperatures, $T_{i}$, and the species flow velocities, $\mathbf{u}_{i}$.
However, these are not associated with any conserved quantity even for a
molecular fluid (there is continual energy and momentum exchange between
species, even though the total energy and momentum are conserved). In this
section, the possibility of including species properties, beyond their
number densities, as hydrodynamic fields is discussed.

The species numbers and flow fields, $n_{i},{\bm u}_{i}$, are related to the
corresponding total fields $n$ and $\mathbf{u}$ by
\begin{equation}
n=n_{1}+n_{2},\quad \rho \mathbf{u=}n_{1}m_{1}\mathbf{u}_{1}\mathbf{+}%
n_{2}m_{2}\mathbf{u}_{2},  \label{4.1}
\end{equation}%
where the mass density is $\rho \equiv m_{1}n_{1}+m_{2}n_{2} = \rho_{1}+ \rho_{2}$. The
temperature is determined from the total energies which are decomposed into
their kinetic temperature and convective energies according to
\begin{equation}
e=\frac{3}{2}nT+\frac{1}{2}\rho u^{2}=e_{1}+e_{2},\hspace{0.25in}e_{i}=\frac{%
3}{2}n_{i}T_{i}+\frac{1}{2} \rho_{i} u_{i}^{2}.  \label{4.1a}
\end{equation}%
Two possible hydrodynamic descriptions are considered here, one for the six
fields $\{n_{i},\mathbf{u},T\}$, and another for the ten fields $\{n_{i},%
\mathbf{u}_{i},T_{i}\}$.

\subsection{Single temperature, single velocity balance equations}

The exact macroscopic balance equations for $n_{1},n_{2},\mathbf{u}$, and $T$
are
\begin{equation}
D_{t}n_{i}+n_{i}\nabla \cdot \mathbf{u}+\nabla \cdot \mathbf{j}_{i}=0,
\label{4.2}
\end{equation}%
\begin{equation}
D_{t}\mathbf{u}+\frac{1}{\rho }\nabla \cdot \mathsf{P}=0,  \label{4.3}
\end{equation}%
\begin{equation}
D_{t}T-\frac{T}{n}\nabla \cdot \sum_{i=1,2}\mathbf{j}_{i}+\frac{2}{3n}\left(
\nabla \cdot \mathbf{q}+\mathsf{P}:\nabla \mathbf{u}\right) =-\zeta T\,,
\label{4.4}
\end{equation}%
where ${\bm j}_{i}$ is the number flow for species $i$ relative to the local
flow, $\mathbf{j}_{i}\equiv n_{i}\left( \mathbf{u}_{i}-\mathbf{u}\right) $.
To obtain hydrodynamic equations for this choice of fields, constitutive
equations must be obtained from the corresponding normal solution to give $%
\mathbf{j}_{i}$, $\mathbf{q}$, $\mathsf{P}$, and $\zeta $ as functionals of
these fields. As with the one component granular fluid, there are two time
scales associated with this choice of hydrodynamic fields, that associated
with the uniformity parameter and that associated with the cooling rate.
Both should be larger than the mean free time.

\subsection{Two temperatures, two velocities balance equations}

The balance equations for the extended set of ten fields have the form
\begin{equation}
D_{it}n_{i}+n_{i}\nabla \cdot \mathbf{u}_{i}=0,  \label{4.5}
\end{equation}%
\begin{equation}
\rho _{i}D_{it}\mathbf{u}_{i}+\nabla \cdot \mathsf{P}_{i}=-{\bm\lambda }_{i}.
\label{4.6}
\end{equation}%
\begin{equation}
D_{it}T_{i}+\frac{2}{3n_{i}}\left( \nabla \cdot \mathbf{q}_{i}+\mathsf{P}%
_{i}:\nabla \mathbf{u}_{i}\right) =-\zeta _{i}T_{i}+\frac{2}{3n_{i}}\mathbf{u%
}_{i}\cdot {\bm\lambda }_{i},  \label{4.7}
\end{equation}%
where now $D_{it}\equiv \partial _{t}+\mathbf{u}_{i}\cdot \nabla $, and the
heat flux $\mathbf{q}_{i}$ and pressure tensor $\mathsf{P}_{i}$ are now
defined as the energy and momentum fluxes in the local rest frame of a fluid
element moving with velocity $\mathbf{u}_{i}$ (compare with Eq.\, (\ref{3.3a})). The
sources ${\bm\lambda }_{i}$ and $\zeta _{i}$ depend on $n_{i}$, $T_{i}$, and
$\mathbf{u}_{i}$. The normal solution for this choice of fields is needed to
get the constitutive equations for $\mathbf{q}_{i}$, $%
\mathsf{P}_{i}$, ${\bm\lambda }_{i}$, and $\zeta _{i}$ as functionals of
these fields.

Note that $\mathbf{q}\neq \mathbf{q}_{1}+\mathbf{q}_{2}$ and $\mathsf{P} \neq \mathsf{P}_{1}+%
\mathsf{P}_{2}$, in general. The balance equations for $n_{1},n_{2},%
\mathbf{u}$, and $T$, Eqs.\ (\ref{4.2})-(\ref{4.4}), follow from Eqs.\ (\ref%
{4.5})-(\ref{4.7}) only if the sources have the properties%
\begin{equation}
\zeta _{1}n_{1}T_{1}+\zeta _{2}n_{2}T_{2}=\zeta nT+\frac{2}{3}\nabla
\cdot \left( \mathbf{s}-\mathbf{s}_{1}-\mathbf{s}_{2}\right) ,
\label{4.8}
\end{equation}%
\begin{equation}
\left( {\bm\lambda }_{1}+{\bm\lambda }_{2}\right) =\nabla \cdot \left(
\mathsf{T-T}_{1}-\mathsf{T}_{2}\right) .  \label{4.8a}
\end{equation}%
Here $\mathbf{s}_{i}$ and $\mathsf{T}_{i}$
are the energy and momentum fluxes of species $i$ in the laboratory frame. These
relations are proved in Appendix \ref{App1}, where explicit
expressions for the sources are obtained.

The coefficients ${\bm\lambda }_{i}$ are proportional to the collision
frequency, and hence introduce dynamics on the short mean free time scale.
The rates $\zeta _{i}$ have two contributions. One is positive and
proportional to the degree of dissipation, like $\zeta $ in Eq.\ (\ref{4.4}%
). However, there is an additional contribution of indefinite sign
and which does not vanish in the elastic limit. It represents the
collisional transfer of energy between different species. This
latter contribution is of the order of the collision rate and
consequently generates dynamics on the mean free time scale. There
are now three hydrodynamic time scales: the largest one set by the
uniformity parameter, that set by the cooling rate, and the fastest
one defined on the mean free time scale. This last time scale is the
central difference between the two hydrodynamic descriptions
being discussed here.

\subsection{Comparison of hydrodynamic descriptions}

To compare and contrast the above two hydrodynamic descriptions, it is
sufficient to consider the homogeneous limit, for which the details of the
constitutive equations are not required. Consider first the case of zero
flow velocities, in which case Eq.\ (\ref{4.1a}) gives%
\begin{equation}
nT=n_{1}T_{1}+n_{2}T_{2}.  \label{4.8b}
\end{equation}%
Equations\ (\ref{4.2})-(\ref{4.4}) reduce to constant species densities, and
a time dependent temperature obtained from
\begin{equation}
\partial _{t}T=-\zeta \left( T\right) T,  \label{4.9}
\end{equation}%
while the two temperature description, Eqs.\, (\ref{4.5})-(\ref{4.7}), gives
also constant species density and %
\begin{equation}
\partial _{t}T_{1}=-\zeta _{1}\left( T_{1},T_{2}\right) T_{1},\hspace{0.25in}%
\partial _{t}T_{2}=-\zeta _{2}\left( T_{1},T_{2}\right) T_{2}.  \label{4.10}
\end{equation}%
Equation (\ref{4.9}) follows from Eqs.\ (\ref{4.10}) via the exact relationships (%
\ref{4.8}) and (\ref{4.8b}) for the uniform case. In this respect, the two
temperature description is more complete. However, it is also more complex
and masks the simplicity of the long time dynamics of (\ref{4.9}) by
superposing on it a higher frequency dynamics, as described in the following.

An approximate representation for the cooling rates is given in Appendix \ref%
{App2} in the form \cite{Santos06}
\begin{equation}
\zeta _{i}=\nu\left( T_{2}\right) \xi _{i}(\phi ),  \label{4.11}
\end{equation}%
where
\begin{equation}
\phi \equiv \frac{m_{2}T_{1}}{m_{1}T_{2}}  \label{4.11a}
\end{equation}%
is a measure of the temperature ratio, and $\nu>0$ is some
average collision frequency. The explicit expressions for $\nu$,
$\xi _{1}$, and $\xi _{2}$
are given in Appendix \ref{App2}. Equations (\ref{4.10}) lead to an equation for $%
\phi $,
\begin{equation}
\partial _{t}\ln \phi =-\nu \left( T_{2}\right) \left( \xi _{1}\left(
\phi \right) -\xi _{2}\left( \phi \right) \right) .  \label{4.12}
\end{equation}%
There is a stationary solution $ \phi_{h}$ determined by%
\begin{equation}
\xi _{1}\left( \phi _{h}\right) =\xi _{2}\left( \phi _{h}\right) .
\label{4.13}
\end{equation}%
It follows from Eqs.\ (\ref{a.2}) and (\ref{a.3}) in Appendix \ref{App2} that $\xi
_{1}\left( \phi \right) -\xi _{2}\left( \phi \right) >0$ for $\phi >\phi
_{h} $ and $\xi _{1}\left( \phi \right) -\xi _{2}\left( \phi \right) <0$ for
$\phi <\phi _{h}$, so the stationary solution is approached in general for
times greater than the mean free time. This solution is the homogeneous
cooling state for a granular mixture.

Stationarity of $\phi $ implies that
the cooling rates of each species are the same,%
\begin{equation*}
\frac{\partial _{t}T_{1}}{T_{1}}=\frac{\partial _{t}T_{2}}{T_{2}}\Rightarrow
\zeta _{1}=\zeta _{2}=\zeta .
\end{equation*}%
The equivalence to $\zeta $ follows from the relationship $nT=n_{1}T_{1}+n_{2}T_{2}$.
Thus all three temperatures are different in general but their cooling rates
are the same. The equivalence of the cooling rates means only one
temperature field is needed to describe the dynamics on this time scale, for
arbitrary initial preparation. For example, in the two temperature case
considered above, the homogeneous solution corresponding to (\ref{3.6}) is
\begin{equation}
\left( -\left( \zeta _{1}-\zeta _{2}\right) \phi \frac{\partial }{\partial
\phi }-\zeta T\frac{\partial }{\partial T}+L\right) \rho _{h}=0,
\label{4.14}
\end{equation}%
where a change of variables has been made from $\{ T_{1},T_{2} \}$ to $\{ T,\phi \}$.
The solution of this equation depends on two time dependent fields, $\rho
_{h}=\rho _{h}(\Gamma ,T(t),\phi (t))$. But on the mean free time scale $%
\phi (t)\rightarrow \phi _{h}$, $\zeta _{1}\left( \phi _{h}\right) =\zeta
_{2}\left( \phi _{h}\right) ,$ and so $\rho _{h}(\Gamma ,T(t),\phi
(t))\rightarrow \rho _{h}(\Gamma ,T(t),\phi _{h})$. The latter is the
homogeneous solution to (\ref{3.6}), the normal solution for the single
temperature case.

This analysis of the homogeneous state illustrates the behavior for
more general inhomogeneous states as well. The slow dynamics on the
scale of the uniformity parameter in (\ref{4.2})-(\ref{4.4}) is
present as well in the two temperature formulation
(\ref{4.5})-(\ref{4.7}), but the latter superposes on that either initial
transients or a continual modulation at the collision frequency.
Since it is only the slow dynamics that is relevant experimentally,
and for justification of the normal solution, the two temperature
formulation is unnecessarily complex. This does not mean that the
slow components of the species temperatures are irrelevant. These
are provided by the single temperature formulation and the normal
solution that gives these species temperatures as functions of the
slow global temperature, $T_{i}(t)=T_{i}(T(t))$. This picture has
been confirmed by molecular dynamics simulations in the homogeneous
case \cite{Dahl} where the rapid approach of $T_{1}/T_{2}$ to a
constant different from unity is observed on the scale of the mean
free time. Similar results have been seen for an inhomogeneous
vibrated mixture of inelastic hard disks \cite{ByR09}, showing that
the two species temperatures in the steady state are determined by
the densities and global temperature profiles, being independent
from the details of the heating mechanism, aside from a boundary
layer next to the vibrating wall.

Finally, consider the species velocities, described by (\ref{4.6}). It is
clear that the source in these equations drives the species velocities
towards the common flow field $\mathbf{u}$, on the mean free time scale.
This is similar to the equilibration of the species temperatures, and is the
same for both molecular and granular systems. As with the species
temperatures, this extended description yields transients and modulation of
the slow dynamics. Instead the single flow hydrodynamics describes only the
relevant slow components of the species flow fields, through the normal
solution, in the form $\mathbf{u}_{i}=\mathbf{u}_{i}(T(t),\mathbf{u}(t))$.

In summary, the two temperature, two velocity hydrodynamic description of a
binary mixture subsumes the one temperature, one velocity formulation by
providing a more detailed description of the fluid. However, the additional
details are on the short time scale of the mean free time where the dynamics
is entangled with many other microscopic degrees of freedom, and for which
the underlying normal solution is not expected to apply. Of course, it has
the additional computational difficulty of requiring the determination of
many more transport coefficients in the constitutive equations \cite{Jenkins}%
, and solving for ten coupled fields in contrast to six fields in the simple
hydrodynamic description.

\section{Discussion}

\label{sec5}The hydrodynamic fields should include all those fields with the
longest time scale, fixed by the uniformity parameter. In addition, fields
with shorter times scales may need to be included. For example, systems with
long internal relaxation times (viscoelastic liquids) can be described by
non-local in time constitutive equations or, equivalently, by local
equations with additional fields. For some non-equilibrium states with
inherent dynamics persisting in the homogeneous state, such as for granular
fluids, non-conserved fields may be essential. Non-conserved species fields
may be useful for cases where their equilibration times in the homogeneous
state are long, such as for electron-ion two temperature plasmas.

A more difficult question is whether to include non-conserved fields whose
time scales include the mean free time, since many other non-hydrodynamic
degrees of freedom are also active on this time scale. There does not seem
to be any formal objection to including such fields, since the
Chapman-Enskog construction of a corresponding normal solution, outlined in
Sec.\, \ref{sec2}, can be implemented. However, the normal solution in such cases is
still restricted to apply only for time scales large compared to the mean
free time. Consequently, the new short time scale component in the
hydrodynamics is not relevant and can obscure the relevant slow dynamics
inherent in all fields.

To make this point in a different way, let $\left\{ \widehat{y}_{\alpha
}\right\} $ be the phase functions whose average values in some
nonequilibrium state are the fields $\left\{ y_{\alpha }\right\} $. It is
possible to write the equations of motion for $\left\{ \widehat{y}_{\alpha
}\right\} $ as  formally exact Langevin equations \cite{Zwanzig,Grabert},
generalizing Eq.\,  (\ref{2.1})%
\begin{equation}
\partial _{t}\widehat{y}_{\alpha }\left( \mathbf{r},t\right) -M_{\alpha }(%
\mathbf{r},t|\left\{ \widehat{y}_{\beta }\right\} )=\widehat{f}_{\alpha
}\left( \mathbf{r},t\right) .  \label{5.1}
\end{equation}%
The functional $M_{\alpha }$ is similar to $N_{\alpha }$ in
(\ref{2.1}), except the former is a time dependent functional,
depending on time
explicitly as well as through its functional argument. The sources $\widehat{%
f}_{\alpha }\left( \mathbf{r},t\right) $ represent the dynamics of all other
degrees of freedom, beyond the set $\left\{ \widehat{y}_{\alpha }\right\} .$
The solution to (\ref{5.1}) is equivalent to that from Newton's equations,
but the dynamics now has been separated into two parts that are supposed to be
distinct: a slow dynamics associated with the homogeneous equation,
modulated by a fast dynamics due to $\widehat{f}_{\alpha }\left( \mathbf{r}%
,t\right) $. The average of these sources vanish, so the average of Eq.\ (\ref{5.1})
resembles a hydrodynamic description,
\begin{equation}
\partial _{t}y_{\alpha }\left( \mathbf{r},t\right) -M_{\alpha }(\mathbf{r}%
,t\mid \left\{ y_{\beta} \right\} )\ldots =0.  \label{5.2}
\end{equation}%
The dots denote additional terms due to fluctuations in the fields, for instance proportional to $%
\left\langle \left( \widehat{y}_{\alpha }-y_{\alpha }\right)
^{2}\right\rangle $, which occur since $M_{\alpha }$ is a nonlinear
functional. Although the sources no longer appear their effects are still
present in the explicit time dependence of $M_{\alpha }$. Only after taking
time scales long compared to that of the sources does this time dependence
become negligible, and a hydrodynamic description obtained. This shows that
general sets of fields can be chosen, but their hydrodynamics is restricted
to long times when a simpler description may be available in terms of fewer
fields.

\section{Acknowledgments}

The research of JJB has been partially supported by the Ministerio de Educaci%
\'{o}n y Ciencia (Spain) through Grant No. FIS2008-01339 (partially financed
by FEDER funds).

\appendix

\section{Momentum and energy sources}

\label{App1}

The sources in the momentum and energy balance equations
are determined in this Appendix, and properties (\ref{4.8}) and
(\ref{4.8a}) are verified. It is assumed that the particles are hard
spheres, although the analysis applies as well for general pairwise
additive forces. The Liouville operator of the system has the form
\begin{equation}
\label{app1.1}
L=\sum_{i=1}^{N_{1}} \frac{{\bm p}_{i1}}{m_{1}} \cdot \frac{\partial}{\partial {\bm q}_{i1}}
+ \sum_{i=1}^{N_{2}} \frac{{\bm p}_{i2}}{m_{2}} \cdot \frac{\partial}{\partial {\bm q}_{i2}}
+ \frac{1}{2} \sum_{j \neq i}^{N_{1}} \sum_{i=1}^{N_{1}} T_{i1j1}+
\frac{1}{2} \sum_{j \neq i}^{N_{2}} \sum_{i=1}^{N_{2}} T_{i2j2}+
\sum_{i=1}^{N_{1}} \sum_{j=1}^{N_{2}} T_{i1j2}\, ,
\end{equation}
where ${\bm q}_{i1}$ (${\bm q}_{i2}$) and ${\bm p}_{i1}$ (${\bm p}_{i2}$) are the
position and momentum of particle $i$ of species $1$ ($2$), and $N_{1}$ and $N_{2}$ are the total number of particles of each species. The operators $T_{ikjl}=T_{jlik}$ describe the
binary collision between particle $i $ of species $k$ and
particle $j $ of species $l$, $k,l=1,2$.

\subsection{Momentum balance equation}

The average momentum density for species $1$ is
\begin{equation}
\mathbf{p}_{1}\left( \mathbf{r}\right) =\left\langle \sum_{i =1}^{N_{1}}%
\mathbf{p}_{i 1}\delta \left( \mathbf{r}-\mathbf{q}_{\alpha 1}\right)
\right\rangle,   \label{b.1}
\end{equation}%
where the brackets denote ensemble average. The time derivative is%
\begin{eqnarray}
\partial _{t}\mathbf{p}_{1}\left( \mathbf{r}\right)  &=&-m_{1}^{-1} {\bm \nabla}
\cdot \left\langle \sum_{i =1}^{N_{1}}\mathbf{p}_{i 1}%
\mathbf{p}_{i 1}\delta \left( \mathbf{r}-\mathbf{q}_{i 1}\right)
\right\rangle +\left\langle \sum_{j \neq i}^{N_{1}} \sum_{i=1}^{N_{1}} T_{i1j1}
\delta \left( \mathbf{r}-\mathbf{q}_{i 1}\right) \mathbf{p}%
_{i 1}\right\rangle   \nonumber \\
&&+\left\langle \sum_{i =1}^{N_{1}}\sum_{j
=1}^{N_{2}}T_{i 1j 2}\delta \left( \mathbf{r}-\mathbf{q}_{i
1}\right) \mathbf{p}_{i 1}\right\rangle .  \label{b.2}
\end{eqnarray}%
The first term on the right hand side of Eq.\ (\ref{b.2}) gives the gradient
of the kinetic part of the momentum flux for species $1$, while the
second term describes collisions among particles of species $1$.
Finally, the last term represents collisions between pairs of
particles of different species. The second term
can be written as the gradient of a flux,
\begin{eqnarray}
\left\langle \sum_{j \neq i}^{N_{1}} \sum_{i=1}^{N_{1}} T_{i1j1}
\delta \left( \mathbf{r}-\mathbf{q}_{i 1}\right) \mathbf{p}%
_{i 1}\right\rangle   &=&\frac{1}{2}\left\langle
\sum_{j \neq i}^{N_{1}} \sum_{i=1}^{N_{1}}
\left( \delta \left( \mathbf{r}-\mathbf{q}_{i 1}\right)
T_{i 1 j 1}\mathbf{p}_{i 1}+\delta \left( \mathbf{r}-\mathbf{q}%
_{j 1}\right) T_{j 1 i 1}\mathbf{p}_{j 1}\right)
\right\rangle   \nonumber \\
&=&\frac{1}{2}\left\langle \sum_{j \neq i}^{N_{1}} \sum_{i=1}^{N_{1}}\left( \delta
\left( \mathbf{r}-\mathbf{q}_{i 1}\right) -\delta \left( \mathbf{r}-%
\mathbf{q}_{j 1}\right) \right) T_{i 1j 1}\mathbf{p}_{i
1}\right\rangle .  \label{b.3}
\end{eqnarray}%
In the last line use has been made of $T_{i 1j 1}\mathbf{p}_{j
1}=-T_{i 1 j 1}\mathbf{p}_{i 1}$ (Newton's third law). Finally,
the difference of the delta functions can be written as a divergence using
the identity%
\begin{eqnarray}
\delta \left( \mathbf{r}-\mathbf{q}_{i 1}\right) -\delta \left( \mathbf{%
r}-\mathbf{q}_{j 1}\right)  &=&\int_{0}^{1}dx\, \frac{\partial}{\partial x}\, \delta \left(
\mathbf{r}-x\mathbf{q}_{i 1}-(1-x)\mathbf{q}_{j 1}\right)   \notag
\\
&=&-{\bm \nabla} \cdot \left( \mathbf{q}_{i 1}-\mathbf{q}_{j
1}\right) \int_{0}^{1}dx\, \delta \left( \mathbf{r}-x\mathbf{q}%
_{i 1}-(1-x)\mathbf{q}_{j1}\right).   \label{b.4}
\end{eqnarray}%
Use of Eq.\, (\ref{b.4}) into Eq.\, (\ref{b.3}) identifies the second term of (\ref{b.2})
as minus the gradient of the momentum flux tensor for species $1$.

A similar analysis of the third term does not lead to a gradient because
exchanging $i ,j $ in the sums leads to two different pairs, for
which Newton's third law does not apply. Hence this last term is the
momentum source in Eq.\ (\ref{4.6}), %
\begin{equation}
{\bm\lambda }_{1}\left( \mathbf{r}\right) =-\left\langle
\sum_{i =1}^{N_{1}}\delta \left( \mathbf{r}-\mathbf{q}_{i
1}\right) {\bm F}_{i 1;2}\right\rangle ,\quad {\bm F}%
_{i 1;2}=\sum_{j =1}^{N_{2}}T_{i 1j  2}\mathbf{p}_{i
1}.  \label{b.5}
\end{equation}%
The property (\ref{4.8a}) follows directly from this result since%
\begin{eqnarray}
{\bm\lambda }_{1}\left( \mathbf{r}\right) +{\bm\lambda }_{2}\left( \mathbf{r}%
\right)  &=&-\left\langle \sum_{i =1}^{N_{1}}\sum_{j
=1}^{N_{2}}\left( \delta \left( \mathbf{r}-\mathbf{q}_{i 1}\right)
T_{i 1 j 2}\mathbf{p}_{i 1}+\delta \left( \mathbf{r}-\mathbf{q}%
_{j 2}\right) T_{j 2 i 1}\mathbf{p}_{j 2}\right)
\right\rangle   \notag \\
&=&-\left\langle \sum_{i =1}^{N_{1}}\sum_{j
=1}^{N_{2}}\left( \delta \left( \mathbf{r}-\mathbf{q}_{i 1}\right)
-\delta \left( \mathbf{r}-\mathbf{q}_{j 2}\right) \right) T_{i
1 j 2}\mathbf{p}_{i 1}\right\rangle .  \label{b.6}
\end{eqnarray}%
where momentum conservation, $T_{i 1 j2}\mathbf{p}_{j
2}=-T_{i 1 j 2}\mathbf{p}_{i 1}$, has been used. Using again
the identity (\ref{b.4}) this becomes the gradient of the momentum flux
tensor due to collisions between particles of different species. This
confirms Eq.\ (\ref{4.8a}).

\subsection{Energy balance equation}

The average energy density for species $1$ is%
\begin{equation}
e_{1}\left( \mathbf{r}\right) =\left\langle \sum_{i=1}^{N_{1}}\frac{%
p_{i 1}^{2}}{2m_{1}}\delta \left( \mathbf{r}-\mathbf{q}_{i
1}\right) \right\rangle\, .   \label{b.7}
\end{equation}%
Its time derivative is given by%
\begin{eqnarray}
\partial _{t}e_{1}\left( \mathbf{r}\right)  &=&-m_{1}^{-1}\nabla \cdot
\left\langle \sum_{i =1}^{N_{1}}\frac{p_{i 1}^{2}}{2m_{1}}\, \mathbf{p%
}_{i 1}\delta \left( \mathbf{r}-\mathbf{q}_{i 1}\right)
\right\rangle +\left\langle
\sum_{j \neq i}^{N_{1}} \sum_{i=1}^{N_{1}}
T_{i 1j 1}\frac{p_{i
1}^{2}}{2m_{1}} \delta \left(
\mathbf{r}-\mathbf{q}_{i 1}\right)\right\rangle   \nonumber \\
&&+\left\langle \sum_{i =1}^{N_{1}}\sum_{j
=1}^{N_{2}} T_{i
1 j 2}\frac{p_{i 1}^{2}}{2m_{1}} \delta \left( \mathbf{r}-\mathbf{q}_{i 1}\right)\right\rangle\, .
\label{b.8}
\end{eqnarray}%
The first term on the right is the divergence of the kinetic part of the
energy flux for species $1$. The second term is due to collisions among
species $1$ particles while the last term is due to collisions between
species $1$ and $2$. The second term can be analyzed as above for the
momentum density,%
\begin{equation*}
\left\langle \sum_{j \neq i}^{N_{1}} \sum_{i=1}^{N_{1}}
T_{i 1j 1} \frac{p_{i 1}^{2}}{%
2m_{1}} \delta \left( \mathbf{r}-%
\mathbf{q}_{i 1}\right) \right\rangle =\frac{1}{2}\left\langle
\sum_{j \neq i}^{N_{1}} \sum_{i=1}^{N_{1}}
\left( \delta \left( \mathbf{r}-\mathbf{q}_{i 1}\right)
T_{i 1 j 1}\frac{p_{i 1}^{2}}{2m_{1}}+\delta \left( \mathbf{r}-%
\mathbf{q}_{j 1}\right) T_{j 1 i 1}\frac{p_{j 1}^{2}}{2m_{1}}%
\right) \right\rangle
\end{equation*}%
\begin{equation}
=-\frac{1}{2}\left\langle
\sum_{j \neq i}^{N_{1}} \sum_{i=1}^{N_{1}}
\delta \left(
\mathbf{r}-\mathbf{q}_{i 1}\right) \Delta _{i 1 j
1}\right\rangle +\frac{1}{2}\left\langle  \sum_{j \neq i}^{N_{1}} \sum_{i=1}^{N_{1}}%
\left[ \left( \delta \left( \mathbf{r}-\mathbf{q}_{j 1}\right) -\delta
\left( \mathbf{r}-\mathbf{q}_{i 1}\right) \right) T_{j 1 i 1}%
\frac{p_{j 1}^{2}}{2m_{1}}\right] \right\rangle \, .  \label{b.9}
\end{equation}%
In the second equality, use has been made again of $T_{i 1 j 1}=T_{j
1 i 1}$ and also of
\begin{equation}
T_{i 1 j 1}\left( \frac{p_{i 1}^{2}}{2m_{1}}+\frac{p_{j
1}^{2}}{2m_{1}}\right) =-\Delta _{i 1 j 1},  \label{b.10}
\end{equation}%
where $\Delta _{i 1 j 1}$ is the energy loss by the pair of particles $i,j$ in the collision. The
difference between the delta functions in the last term of Eq.\ (\ref{b.9})  can be transformed to the divergence
of the energy flux using again Eq.\, (\ref{b.4}). Therefore, the second term on the
right side of (\ref{b.8}) is the sum of the energy loss due to inelasticity
and the energy flux for particles of species $1$ alone.

The last term of (\ref{b.8}) cannot be represented as the divergence of a
flux since interchanging $i $ and $j $ in the summation introduces a
different pair. The total energy source for species $1$ is then identified as%
\begin{equation}
w_{1}(\mathbf{r}) \equiv - \frac{1}{2}\left\langle \sum_{j \neq i}^{N_{1}} \sum_{i=1}^{N_{1}} \delta \left( \mathbf{r}-\mathbf{q}_{i 1}\right) \Delta
_{i 1 j 1}\right\rangle + \left\langle \sum_{i
=1}^{N_{1}}\delta \left( \mathbf{r}-\mathbf{q}_{i 1}\right) G
_{i 1;2}\right\rangle ,\quad G_{i
1;2}=\sum_{j =1}^{N_{2}}T_{i 1j 2}\frac{p_{i 1}^{2}}{2m_{1}%
}\, .  \label{b.11}
\end{equation}%
The cooling rates in the temperature equation are proportional to $w_{i},(%
\mathbf{r})$%
\begin{equation}
\zeta _{1}T_{1}=- \frac{2}{3n_{1}}w_{1}.  \label{b.12}
\end{equation}%

The sum of the two sources is%
\begin{equation*}
w_{1}(\mathbf{r})+w_{2}(\mathbf{r})=- \frac{1}{2}\left\langle
\sum_{j \neq i}^{N_{1}} \sum_{i=1}^{N_{1}}
\delta \left( \mathbf{r}-\mathbf{q}_{i 1}\right)
\Delta _{i 1 j 1}+ \sum_{j \neq i}^{N_{2}} \sum_{i=1}^{N_{2}}
\delta \left(
\mathbf{r}-\mathbf{q}_{j 2}\right) \Delta _{i 2 j
2}\right\rangle
\end{equation*}%
\begin{equation}
+ \left\langle \sum_{i =1}^{N_{1}}\sum_{j
=1}^{N_{2}}\left( \delta \left( \mathbf{r}-\mathbf{q}_{i 1}\right)
T_{i 1 j 2}\frac{p_{i 1}^{2}}{2m_{1}}+\delta \left( \mathbf{r}-%
\mathbf{q}_{j 2}\right) T_{j 2 i 1}\frac{p_{j 2}^{2}}{%
2m_{2}}\right) \right\rangle .  \label{b.13}
\end{equation}%
Using (\ref{b.10}) this becomes
\begin{eqnarray}
w_{1}(\mathbf{r})+w_{2}(\mathbf{r}) & = &- \left\langle
\frac{1}{2} \sum_{j \neq i}^{N_{1}} \sum_{i=1}^{N_{1}}
\delta \left( \mathbf{r}-\mathbf{q}_{j 1}\right)
\Delta _{i 1 j 1}+ \frac{1}{2} \sum_{j \neq i}^{N_{2}} \sum_{i=1}^{N_{2}}
\delta \left(
\mathbf{r}-\mathbf{q}_{j 2}\right) \Delta _{i 2 j
2} \right. \nonumber \\
&& \left. +\sum_{i =1}^{N_{1}}\sum_{j=1}^{N_{2}}\delta \left( \mathbf{r}-%
\mathbf{q}_{j 2}\right) \Delta _{i1j2}\right\rangle \nonumber \\
& & + \left\langle \sum_{i =1}^{N_{1}}\sum_{j
=1}^{N_{2}}\left( \delta \left( \mathbf{r}-\mathbf{q}_{i 1}\right)
-\delta \left( \mathbf{r}-\mathbf{q}_{j 2}\right) \right) T_{i
1 j 2}\frac{p_{i 1}^{2}}{2m_{1}}\right\rangle .  \label{b.14}
\end{eqnarray}
The first term is the total energy loss due to all inelastic collisions,
while the last term becomes (using (\ref{b.4})) the contribution to the
divergence of the energy flux from collisions between particles of different
species. This confirms Eq.\ (\ref{4.8}).

\section{Energy exchange rates}

\label{App2}The "cooling" rates for a binary mixture, $\zeta _{i}$, appearing in (\ref%
{4.10}) have been estimated from an approximate two particle reduced
distribution function (local equilibrium or information entropy
distribution), to give their dependence on the fields $n_{i},T_{i}$ and the
mechanical properties of the particles. It is assumed that the particles are
hard, smooth, inelastic spheres with masses, diameters, and densities $%
m_{i},\sigma _{i}$, and $n_{i}$ respectively. The inelasticity is measured
by restitution coefficients $\alpha _{ij}$ for collisions between \
particles of type $i$ and $j$, with $0<\alpha _{ij}\leq 1$ $\ $and $\alpha
_{ij}=1$ representing elastic collisions. The cooling rates are proportional
to an average collision frequency $\nu $ with the approximate results
\cite{Santos06}
\begin{equation}
\zeta _{i}=\nu \xi _{i},\quad \nu =\frac{4\pi m_{2}}{3m}\chi
_{21}n\sigma _{21}^{2}\sqrt{\frac{8T_{2}}{\pi m_{2}}}\left( 1+\alpha
_{21}\right) ,  \label{a.1}
\end{equation}%
\begin{equation}
\xi _{1}=(1-\alpha _{11}^{2})x_{1}\beta _{1}\sqrt{\phi }+x_{2}\frac{\sqrt{%
1+\phi }}{1+\phi _{0}}\left( 1-\mu +\phi _{0}-\frac{\mu }{\phi }\right) ,
\label{a.2}
\end{equation}%
\begin{equation}
\xi _{2}=\left( 1-\alpha _{22}^{2}\right) x_{2}\beta _{2}+x_{1}\frac{\sqrt{%
1+\phi }}{1+\phi _{0}}\left( 1+\frac{\phi _{0}-\phi }{\mu }+\phi \right)
(1-\mu ).  \label{a.3}
\end{equation}%
The variable $\phi $ is a measure of the temperature ratio defined in Eq.\ (\ref{4.11a}), $\sigma
_{21}\equiv (\sigma _{1}+\sigma _{2})/2$, $m ºequiv m_{1}+m_{2}$,  $\chi_{ij}$ is the pair correlation function at contact
of a particle of species $i$ and a particle of species $j$, $x_{i}=n_{i}/n$,  and
the constants $\beta _{1}, \beta_{2}, \phi _{0},$ and $\mu $ are given by
\begin{equation}
\beta _{1}=\frac{1+\phi _{0}}{4\sqrt{2}\mu }\frac{\chi _{11}}{\chi _{21}}%
\left( \frac{\sigma _{1}}{\sigma _{21}}\right) ^{2},\quad \beta _{2}=\frac{%
1+\phi _{0}}{4\sqrt{2}\mu }\frac{\chi _{22}}{\chi _{21}}\left( \frac{\sigma
_{2}}{\sigma _{21}}\right) ^{2}\,,  \label{a.4}
\end{equation}%
\begin{equation}
\phi _{0}=\frac{1-\alpha _{21}}{1+\alpha _{21}},\quad \mu =\frac{m_{2}}{m}\,.
\label{a.5}
\end{equation}%
As noted in the text, $\zeta _{i}$ has two contributions, one that is
positive and vanishes as $\left( 1-\alpha _{ii}^{2}\right) $ in the elastic
limit, and another that remains finite in that limit.

\end{document}